\documentclass[preprint,12pt]{elsarticle}

\usepackage{amssymb}
\usepackage{amsthm}
\usepackage{mathtools, nccmath}
\usepackage{lipsum}
\usepackage{multicol}
\usepackage{natbib}
\usepackage{graphicx} 
\usepackage{subcaption}
\usepackage{amsmath}
\usepackage{color}
\usepackage{bbm}
\biboptions{sort&compress}
\usepackage[T1]{fontenc}

\journal{Computer Physics Communications}

\begin{document}

\begin{frontmatter}



\title{Catalytic flow with a coupled Finite Difference - Lattice Boltzmann scheme}


\author[HIERN]{Nadiia Kulyk\corref{cor1}}
\cortext[cor1]{The authors contributed equally to this work}
\address[HIERN]{Helmholtz Institute Erlangen-N\"urnberg for Renewable Energy (IEK-11), Forschungszentrum J\"ulich, F\"urther Stra\ss e 248, 90429 N\"urnberg, Germany}

\author[HIERN]{Daniel Berger\corref{cor1}}

\author[PULS,Croatia]{Ana-Sun\v{c}ana Smith}

\address[PULS]{PULS Group, Department of Physics, Interdisciplinary Center for Nanostructured Films, Friedrich-Alexander-University of Erlangen-N{\"u}rnberg, Cauerstra\ss e 3, 91058 Erlangen, Germany}

\address[Croatia]{Group for Computational Life Sciences, Division of Physical Chemistry, Ru\dj er Bo\v{s}kovi\'c Institute,
 Bijeni\v{c}ka cesta 54, P.P. 180, HR-10002 Zagreb,
Croatia}

\author[HIERN,Eindhoven]{Jens Harting}

\address[Eindhoven]{Department of Applied Physics, Eindhoven University of Technology, P. O. Box 513, 5600MB Eindhoven, The Netherlands }

\begin{abstract}
Many catalyst devices employ flow through porous structures, which leads to a complex macroscopic mass and heat transport. To unravel the detailed dynamics of the reactive gas flow, we present an all-encompassing model, consisting of thermal lattice Boltzmann model by Kang et al., used to solve the heat and mass transport in the gas domain, coupled to a finite differences solver for the heat equation in the solid \textit{via} thermal reactive boundary conditions for a consistent treatment of the reaction enthalpy. The chemical surface reactions are incorporated in a flexible fashion through flux boundary conditions at the gas-solid interface. We scrutinize the thermal FD-LBM by benchmarking the macroscopic transport in the gas domain as well as conservation of the enthalpy  across the solid-gas interface. We exemplify the applicability of our model by simulating the reactive gas flow through a microporous material catalysing the so-called water-gas-shift reaction.

\end{abstract}



\begin{keyword}
Catalytic flow \sep thermal lattice Boltzmann method \sep reaction enthalpy \sep conjugated heat transfer


\end{keyword}

\end{frontmatter}


\section{Introduction}
Progressing towards a sustainable and clean energy future is one of the main challenges of the 21$^{st}$ century. In this context, catalysis plays a key role in the production of chemical energy storage, efficient chemical conversion, and removal of air pollution. 
Many catalytic devices employ porous structures to maximize the surface and thus maximize the conversion rate.
Such structures in turn induce complex mass transport by the gas flow and even more complex heat transport phenomena through the gas and the solid part of the reactor device, where local heat build-up compromises the reactor stability~\cite{doi:10.1021/cr960406n,B807080F,mehnert2005supported}.
While experimental insight is often not possible at the necessary resolution, computer simulations offer unique and detailed information over a wide range of length and time scales. 

Over the last decades the lattice Boltzmann method (LBM)~\cite{benzi1992lattice,kruger2017lattice} has become a reliable and fast technique for the simulation of complex flows at mesoscopic length scales, such as in many porous materials~\cite{liu2016multiphase}.
In contrast to solving the Navier-Stokes equation, the LBM incorporates microscopic information in the form of single-particle distribution functions as used in the Boltzmann equation, and it is predestined for application on high performance computers due to its straightforward parallelization.
While the traditional LBM is a single component athermal method, several approaches have been suggested to include temperature dynamics. The most popular are the two-population approach, where a separate population function is used to describe temperature evolution~\cite{PhysRevE.55.2780, he1998novel,PhysRevE.68.026701,PhysRevE.75.036704}, the multi-speed LBM, where a larger number of discrete velocities is used~\cite{scagliarini2010lattice}, and the correction terms model~\cite{PhysRevE.87.053304,PhysRevE.89.063310}. The simulations were also extended to include dynamics of arbitrary gas-mixtures~\cite{Shan1995,shan1996diffusion,PhysRevE.76.046703,liu2016multiphase,HKH11}.

Kang \textit{et al.} \cite{PhysRevE.87.053304,PhysRevE.89.063310} suggested a multicomponent thermal LBM. In this model, each chemical gas component is described by individual distribution functions and relaxation parameters modelling the individual transport coefficients (viscosity, thermal and mass diffusivity), where self-consistent correction terms restore the transport equations in the macroscopic limit. 
Surface chemical reactions can hereby be included through flux boundary conditions in a very flexible way~\cite{PhysRevE.78.046711,PhysRevE.89.063310}. This approach however mimics isothermal walls. Therefore, it lacks the ability to simulate a heat built-up, and does not allow for heat transport across the solid-gas interface and through the solid.
In this paper we extend the multicomponent thermal LBM scheme by Kang \textit{et al.} by augmenting it with a finite difference (FD) solver for the heat equation in the solid parts of the simulation domain (e.g. the porous host material), with a special focus on a coupling strategy for consistent treatment of the reaction enthalpy.
Although various approaches exist to address conjugated heat transfer \cite{WANG2007228,CHEN2007266,MENG20081203,CHEN201383,PhysRevE.88.063310,PhysRevE.89.043308,PhysRevE.91.033306,PhysRevE.92.063305,PhysRevE.91.023304,RIHAB2016728, CHEN2017862,PhysRevE.98.043309,doi:10.1080/10407782.2018.1444868}, coupling to a consistent thermal multicomponent model including the reaction enthalpy is a novelty.

The article is structured as follows. In the following section, the thermal multicomponent lattice Boltzmann model based on Ref.~\cite{PhysRevE.89.063310} is introduced. Then, the macroscopic limit of mass and energy transport of the model is shown and we detail our implementation.
We scrutinize our model by benchmarking the mass and thermal diffusivities in the multicomponent system and furthermore simulate a thermal Couette flow for a mixture of gases. In the section on the finite-difference lattice Boltzmann scheme, we present our model to simulate thermal diffusion through a solid-fluid interface by adding a finite difference (FD) solver for the heat equation in the solid.
We exemplify our model with an enthalpy consistent simulation of a reactive flow through a packed-cubes model geometry. The surface of the cubes catalyzes the water-gas-shift reaction, the catalytic reaction of carbon monoxide and water to carbon dioxide and hydrogen. This reaction is of high relevance for the production of highly purified hydrogen gas for application in fuel cells.
The paper closes with our conclusions and a discussion on the limitations of the presented model in resolution and parameter space.

\section{Thermal multicomponent Model}\label{sec:thermalmulticomponentmodel}
\begin{figure}
  \centerline{\includegraphics[width=0.5\columnwidth]{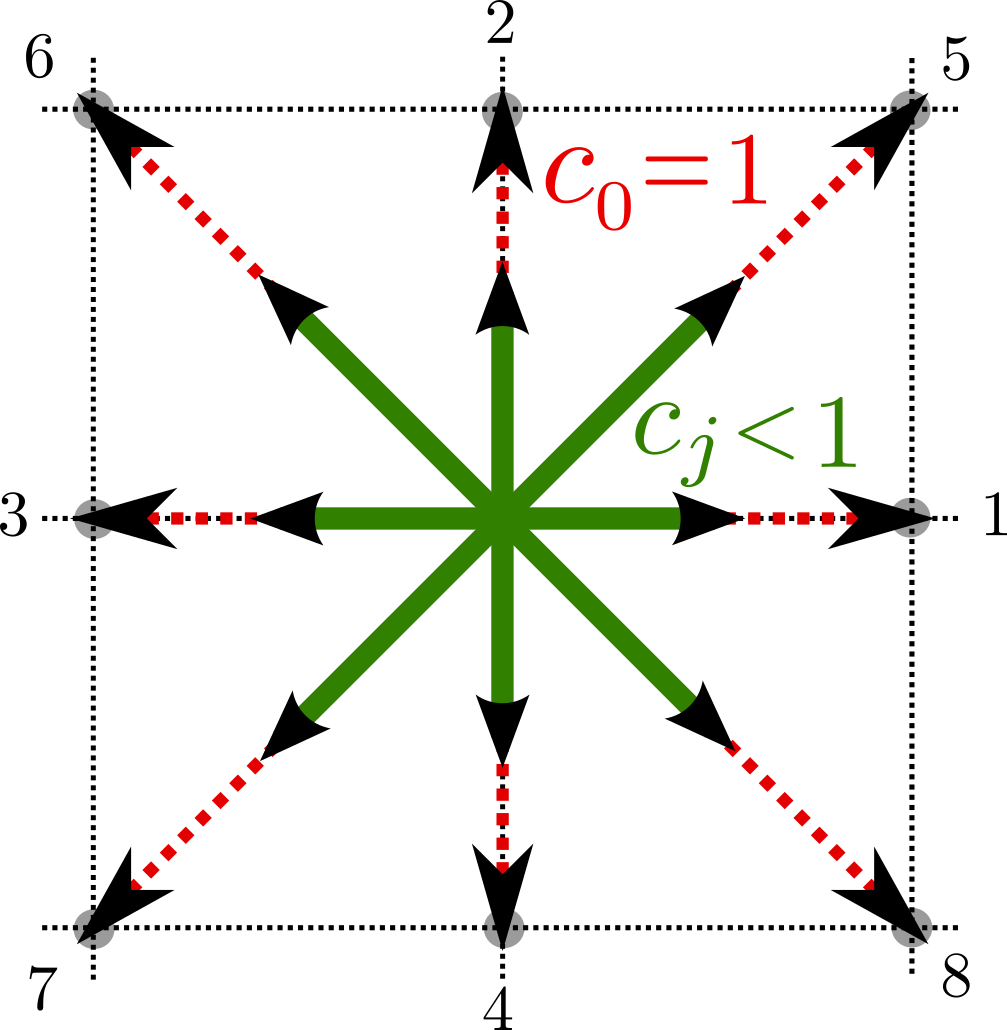}}
 \caption{\label{fig:D2Q9} The D2Q9 lattice: only the lightest element with $c_j=1$ (red) is streamed on-lattice, while all heavier elements are streamed off-lattice with $c_j<1$ (green).
 }
\end{figure}

In this section the thermal multicomponent LBM based on the work of Kang \textit{et al.}  \cite{PhysRevE.87.053304,PhysRevE.89.063310} is introduced shortly. We restrict ourselves to the standard D2Q9-lattice (see Fig.~\ref{fig:D2Q9}) \cite{qian1992lattice}. An extension to the general 3-dimensional case is planned for the future.
In the thermal multicomponent LBM the distribution functions $f_j$ of each component $j$ of a mixture of $N$ components are propagated by a kinetic equation involving advection and collision
\begin{equation} \label{eq:bare_kinetic_eq}
\partial_t f_{ji} + c_{ji\alpha} \partial_\alpha f_{ji} = - \frac{1}{\tau_{1j} } (f_{ji} - f^{*}_{ji}) - \frac{1}{\tau_{2j} } (f^{*}_{ji} - f^{eq}_{ji}) ,
\end{equation}
where $f^{eq}_{ji}$ is the equilibrium distribution function, $f^{*}_{ji}$ is the distribution function of an auxiliary quasi-equilibrium state and
$i$ indices the microscopic velocities $c_{ji\alpha}$, with $\alpha$ being the direction along the x and y axis. On the D2Q9 lattice the microscopic velocities $c_{ji\alpha}$ then read
\begin{equation}
\begin{aligned}
\label{eq:cx}
c_{jix}  &= c_j (0,1,0,-1,0,1,-1,-1,1 );\\
c_{jiy}  &= c_j (0,0,1,0,-1,1,1,-1,-1) ,
\end{aligned}
\end{equation}
which corresponds to a lattice spacing of $\delta x = 1$ and a time step of $\delta t = 1$.
In general, each component has a different mass and thus also different microscopic velocities. This manifests in two ways: first, $c_{ji}$ are scaled with a factor $c_j=\sqrt{M_0/M_j}$ (for convenience we scale all molar masses relative to the mass of the lightest element $M_j \longrightarrow M_j/M_0$ from here on).
Second, the streaming step is impeded which will be discussed further below.
In contrast to the single component LBM the collision operator in Eq.~(\ref{eq:bare_kinetic_eq}) separates into two terms:
based on a fast--slow decomposition along a quasi--equilibrium trajectory \cite{GORBAN1994401,Levermore1996} the multicomponent Bhatnagar-Gross-Krook~\cite{bhatnagar1954model} collision operator relaxes the distribution towards the mixture equilibrium distribution $f^{eq}_{ji}$ in a two-step process via the auxiliary equilibrium for each component $f^{*}_{ji}$ as motivated by Arcidiacono \textit{et al.} \cite{PhysRevE.76.046703} and Ansumali \textit{et al.} \cite{ansumali2007quasi}.

The relevant moments of the $f_{ji}$, which are needed for the analysis of the model, are the individual density, momentum, energy, pressure tensor, third order moment, energy flux and the individual fourth order contracted moment:
\begin{equation} \label{eq:moments_of_f_j}
\begin{split}
\rho_j &= \sum_{i=0}^{8} f_{ji}, \;\;\;\; 
J_{j\alpha} = \sum_{i=0}^{8} c_{ji\alpha} f_{ji},  \;\;\;\;  E_{j} = \sum_{i=0}^{8} c^2_{ji} f_{ji}, \\ 
P_{j\alpha\beta}  &= \sum_{i=0}^{8} c_{ji\alpha} c_{ji\beta} f_{ji},  \;\;\;\;  Q_{j\alpha\beta\gamma} = \sum_{i=0}^{8} c_{ji\alpha}c_{ji\beta}c_{ji\gamma} f_{ji},  \\
q_{j\alpha} &= \sum_{i=0}^{8} c_{ji\alpha} c^2_{ji} f_{ji},  \;\;\;\;  R_{j\alpha\beta} = \sum_{i=0}^{8} c_{ji\alpha}c_{ji\beta}c^2_{ji} f_{ji}. \\
\end{split}
\end{equation}

The equilibrium distribution function $f^{eq}_{ji}$ is found from minimizing the $H$-theorem ($H = \sum_{ji} f_{ji} \ln f_{ji}$) under the constraints of local conservation of the individual density, the mixture momentum, mixture pressure tensor and energy~\cite{PhysRevE.76.016702,PhysRevE.87.053304,PhysRevE.89.063310}. In the specific case of the D2Q9-lattice $f^{eq}_{ji}$ can be written as
\begin{equation}
\label{eq:eq_state_pop}
f^{eq}_{ji} = \rho_j \prod_{\alpha=x,y} \frac{1-2c^2_{0i\alpha}}{2^{c^2_{0i\alpha}}} [(c^2_{0i\alpha}-1)
+ \sqrt{M_j}{c}_{0i\alpha}\frac{J_{\alpha}}{\rho} + M_j\frac{J^2_{\alpha}}{\rho^2} + T] ,
\end{equation}
where $c_{0i\alpha}$ are the microscopic velocities of the lightest element ($M_0=1$).
The equilibrium is defined by the individual densities $\rho_j$, the momentum $J_\alpha$ and the temperature $T$ of the mixture, which is defined by the total energy of the system: 
\begin{equation} \label{eq:moments_of_f}
\begin{split}
\rho &= \sum_{j}^{N} \rho_j = \sum_{j}^{N} \sum_{i=0}^{8} f_{ji}, \\
J_{\alpha} &= \sum_{j}^{N} J_{j\alpha} = \sum_{j}^{N} \sum_{i=0}^{8} c_{ji\alpha} f_{ji}, \\
E &= 2CT + \frac{J^2}{\rho} = \sum_{j}^{N} E_{j} = \sum_{j}^{N} \sum_{i=0}^{8}c^2_{ji} f_{ji}.
\end{split}
\end{equation}
The concentration of the mixture $C$ is the sum of the individual concentrations $C = \sum_{j}^{N} C_j$, where $C_j = \rho_j/M_j$.
The distribution of the auxiliary state can be derived in a similar fashion as the $f^{eq}_{ji}$~\cite{PhysRevE.78.046711}.
For flow simulations of gas mixtures in a porous material, the local mass transport is typically governed by diffusion, i.e.~the individual momentum is the variable which only slowly relaxes towards the equilibrium and is thus conserved in the quasi-equilibrium state. In this case the form of $f^{*}_{ji}$ is very similar to $f^{eq}_{ji}$, with the difference that the momentum of the mixture is replaced by the individual moments,
\begin{equation}
\label{eq:aux_state_pop}
f^{*}_{ji} =
\rho_j \prod_{\alpha=x,y} \frac{1-2c^2_{0i\alpha}}{2^{c^2_{0i\alpha}}} [(c^2_{0i\alpha}-1)
+ \sqrt{M_j}{c}_{0i\alpha}\frac{J_{j\alpha}}{\rho_j} + M_j\frac{J^2_{j\alpha}}{\rho^2_j} + T]. 
\end{equation}
Note that for the case that the relaxation towards the equilibrium is governed by viscosity, i.e. the individual pressure difference equilibrates slowly, a different form of $f^{*}_{ji}$ has to be derived~\cite{PhysRevE.78.046711}.

The moments of the equilibrium distribution recover those from the kinetic theory of gases up to the second order by construction. Deviations however appear for the higher moments, and are labelled $Q'_{j\alpha\beta\gamma}$, $q'_{j\alpha}$ and $R'_{j\alpha\beta}$. These deviation terms stem from the simplicity of the lattice itself. In principle, an equilibrium distribution which also conserves higher moments can be derived for a lattice which includes more velocities~\cite{0295-5075-42-4-359,PhysRevLett.97.190601,rubinstein2008theory}. This strategy however quickly becomes cumbersome, as e.g. the computation of the pressure tensor scales quadratically with the number of velocities ($\propto \mathcal{O}(N^2)$). Here, we choose the strategy of Kang et al. \cite{PhysRevE.87.053304} by explicitly calculating the deviation from the macroscopic limits by means of a Chapman-Enskog analysis~\cite{chapman1970mathematical}. The deviation terms can then be corrected for by adding appropriate forcing terms to the kinetic equation of the model (see next chapter).

\section{Macroscopic limit}\label{sec:macroscopiclimit}
\subsection{Mass diffusivity}
Following Arcidiacono et al.~\cite{PhysRevE.78.046711},
\begin{equation} \label{eq:tau2_approx_text}
\tau_{j2} = \frac{\rho_j }{X_j p}  D_j,
\end{equation}
with 
\begin{equation} \label{eq:D_j_text}
 D_j = \frac{1-Y_j}{ \sum_{\substack{k=1 \\ j\neq k}}^N\frac{ X_k}{D_{jk}} }
\end{equation}
being the mixture averaged diffusion coefficient of species $j$.
This approximation of mass-averaged diffusion velocities impairs the total momentum conservation 
\begin{equation} \label{eq:transport_total_mom}
\partial_t J_{\alpha} + \partial_\beta P_{\alpha\beta}= -\sum_j \frac{V_{j\alpha}}{\tau_{j2}} = - \sum_j \frac{1}{\tau_{j2}} \bigg(J_{j\alpha} - \frac{\rho_j}{\rho}J_{\alpha} \bigg),
\end{equation}
where the terms on the right hand side generally do not add up to zero, and hence act as a spurious force on the gas-mixture. We restore the total momentum conservation by adding a corrective diffusion velocity $U^c_{\alpha}$ \cite{oran1981detailed,coffee1981transport}
to the individual mass flux
\begin{equation} \label{eq:correction_diffusion_velocities}
J_{j\alpha} = \widetilde{J}_{j\alpha} + \rho_j U^c_{\alpha},
\end{equation}
where $\widetilde{J}_{j\alpha}$ is the uncorrected momentum of species $j$. This technique has also been applied in previous multicomponent lattice Boltzmann approaches \cite{PhysRevE.76.046703,PhysRevE.89.063310}. 
Inserting Eq.~(\ref{eq:correction_diffusion_velocities}) into Eq.~(\ref{eq:transport_total_mom}) yields a final expression for the diffusion velocity correction
\begin{equation} \label{eq:correction_diffusion_velocities_final}
U^c_{\alpha} = \frac{\sum_j \frac{1}{\tau_{j2}} \bigg(\widetilde{J}_{j\alpha} - \frac{\rho_j}{\rho} J_{\alpha} \bigg)}{\sum_j \frac{1}{\tau_{j2}} \rho_j}.
\end{equation}
This velocity correction is then added to the kinetic equation (Eq.~(\ref{eq:bare_kinetic_eq})) using a forcing term
\begin{equation} \label{eq:forcing_correction_diffusion_velocities}
\Psi^{(I)}_{ji} = \psi_{ji\alpha} \frac{\rho_j U^c_{\alpha}}{\tau_{j2}},
\end{equation}
where the matrix coefficient $\psi_{ji\alpha}$ is chosen in a way that $\Psi^{(I)}_{ji}$ solely affects the momentum equation:
\begin{equation} \label{eq:psi_jialpha}
\begin{split} 
\psi_{jix} &= \frac{1}{4c_j} \{0,4,0,-4,0,-1,1,1,-1\} \\
\psi_{jiy} &= \frac{1}{4c_j} \{0,0,4,0,-4,-1,-1,1,1\} 
 \end{split}
\end{equation}
The velocity correction discussed above is only a first-order approximation to the exact solution. Higher orders were discussed e.g. by Oran and Boris \cite{oran1981detailed}, but go beyond the scope of this implementation.

\subsection{Viscosity}\label{subsec:viscosity}

There is no trivial way to express the viscosity of a mixture by the viscosity of the individual species $\mu_j$. The empirical formula by Wilke \cite{wilke1950viscosity,welty2009fundamentals} can be used to express the viscosity of the mixture by the viscosities, molar fractions and molar weights of the individual components,
\begin{equation} \label{eq:mu_mixture_wilke}
\mu =  \sum^N_j  \frac{X_j \mu_j}{\sum^N_k X_k \varphi_{jk}} ,
\end{equation}
with
\begin{equation} \label{eq:wilke}
\varphi_{jk} = \frac{1}{\sqrt{8}} \frac{1}{\sqrt{1+\frac{M_j}{M_k}}} \bigg[1 + \bigg(\frac{\mu_j}{\mu_k}\bigg)^{1/2} \bigg(\frac{M_k}{M_j}\bigg)^{1/4}\bigg]. 
\end{equation}
The dynamic viscosity of a mixture in the thermal LBM is~\cite{PhysRevE.89.063310}
\begin{equation} \label{eq:multicomponent_mu}
\mu =  \sum^N_j  (\tau_{j1} C_j T) .
\end{equation}
Combining Eq.~(\ref{eq:multicomponent_mu}) and the Wilke formula allows to express the relaxation times $\tau_{j1}$ as a function of the individual dynamic viscosities, molar fractions and molar masses as
\begin{equation} \label{eq:tau1s}
\tau_{j1} =  \frac{\mu_j}{C T \sum^N_k X_k \varphi_{jk}} .
\end{equation}

\subsection{Heat conductivity}
The thermal conductivity of the mixture is determined as
\begin{equation} \label{eq:multicomponent_kappa}
\kappa =  \bar{M}\sum_{j}  2\tau_{j1}\frac{C_j T}{M_j} ,
\end{equation}
with $\bar{M}$ being the average mass, as previously derived by Kang et al.~\cite{PhysRevE.89.063310}.

One important measure of a gas is its ratio between viscous and thermal diffusivities, which is known as the Prandtl number ($Pr = \frac{c_p \mu}{\kappa}$). 
Similar to Ref.~\cite{PhysRevE.76.016702} a variable Prandtl number can be introduced by adding a term 
\begin{equation} \label{eq:forcing_correction_Pr}
\Phi^{Pr}_{ji} = \phi_{ji} \partial_\alpha q^{Pr}_{j\alpha}
\end{equation}
to the kinetic equation, where 
\begin{equation} \label{eq:q_Pr}
q^{Pr}_{j\alpha} = \bigg(\frac{2c_p}{Pr} - \frac{4}{M_j}\bigg) \tau_{1j} C_jT \partial_\alpha T 
\end{equation}
is ensured to only act on the energy equation through the coefficients $\phi_{ji}$.
The second term compensates the first term of the nonequilibrium energy flux, while the first term sets the thermal conductivity to
\begin{equation} \label{eq:multicomponent_kappa_Pr}
\kappa^{Pr} =  \frac{c_p \sum_j \tau_{j1}C_j T}{Pr}  \;.
\end{equation}
Further extensions in which the Prandtl number can be adjusted for each species separately are possible~\cite{PhysRevE.89.063310}.

\section{Implementation}\label{sec:implementation}
\subsection{Corrected LBM}
In order to achieve that the macroscopic mass and energy transport is in agreement with the kinetic theory of gases, the correction terms from the previous section need to be added to the kinetic equation
\begin{equation} \label{eq:corrected_kinetic_eq}
\partial_t f_{ji} + c_{ji\alpha} \partial_\alpha f_{ji} =
- \frac{1}{\tau_{1j} } (f_{ji} - f^{*}_{ji}) - \frac{1}{\tau_{2j} } (f^{*}_{ji} - f^{eq}_{ji}) 
 + \Psi_{ji} + \Phi_{ji}  .
\end{equation}
Eq.~(\ref{eq:corrected_kinetic_eq}) is now the kinetic equation for a thermal flow of a general gas mixture, which conserves momentum and energy. This equation is to be integrated in time from $t$ to $t+\delta t$. In order to avoid implicitness the distributions $f_{ji}$ are transformed in standard fashion to
\begin{equation} \label{eq:trafo_f_to_g}
g_{ji} =
f_{ji} + \frac{\delta t}{2 \tau_{1j} } (f_{ji} - f^{*}_{ji}) + \frac{\delta t}{2\tau_{2j} } (f^{*}_{ji} - f^{eq}_{ji}) 
- \frac{\delta t}{2} (\Psi_{ji} + \Phi_{ji} ) 
\end{equation}
as done in previous approaches~\cite{he1998novel,ansumali2007quasi,PhysRevE.76.016702,PhysRevE.76.046703,PhysRevE.89.063310}.
Time integration through the trapezoidal rule yields the equation
\begin{equation} \label{eq:final_kinetic_equation}
\begin{split} 
g_{ji}(t+\delta t) =& g_{ji}(t) -  \frac{2\delta t}{\delta t + 2 \tau_{1j} } [g_{ji}(t) - f^{*}_{ji}(t) ] \\
&- \frac{2\delta t}{\delta t+2\tau_{1j} } \frac{\tau_{j1}}{\tau_{j2}}[f^{*}_{ji}(t)  - f^{eq}_{ji}(t) ] \\
 &+ \frac{2 \delta t \tau_{j1}}{\delta t+2\tau_{1j} }  [\Psi_{ji}(t)  + \Phi_{ji}(t)] ,
 \end{split}
\end{equation}
which relates the pre-collision population $g_{ji}(t)$ to the post-collision population $g_{ji}(t+\delta t)$.
The functions $f^{*}_{ji}$, $f^{eq}_{ji}$, $\Phi_{ji}$ and $\Psi_{ji}$ require the moments of the distribution function $f_{ji}$ (Eq.~(\ref{eq:moments_of_f_j})). 
We can, however, calculate the relevant moments of $f_{ji}$ from the populations $g_{ji}$ following the instructions in \cite{ansumali2007quasi,PhysRevE.89.063310}:
\begin{equation} \label{eq:moments_of_g}
\begin{split}
\rho_j(f) &= \rho_j(g) = \sum_{i} g_{ji}, \\
J_{j\alpha}(f) &= \frac{J_{j\alpha}(g) + \frac{\delta t}{2\tau_{j2}} J_{j\alpha}(f^{eq}) +\frac{\delta t}{2} \sum_i c_{ji\alpha}\Psi_{ji}}{1+ \frac{\delta t}{2\tau_{j2}}}, \\
 &= \frac{\sum_i c_{ji\alpha}g_{ji} + \frac{\delta t}{2\tau_{j2}} \sum_i c_{ji\alpha}f^{eq}_{ji}+\frac{\delta t}{2} \sum_i c_{ji\alpha}\Psi_{ji}}{1+ \frac{\delta t}{2\tau_{j2}}}, \\
T(f) &= \frac{1}{2C} \bigg[ E(g) - \frac{(J(g))^2}{\rho} \bigg] + \frac{\delta t}{4C} \sum_j \sum_i c_{ji}^2 \Phi_{ji}.
\end{split}
\end{equation}
This, of course, has the immediate advantage that the populations $f_{ji}$ are no longer needed during the simulations, and transformations between $f_{ji}$ and $g_{ji}$ can be avoided.

The correction terms $\Phi$ and $\Psi$ involve the calculation of gradients, which are all obtained through a second-order finite difference scheme.

The relaxation times $\tau_{1j}$ and $\tau_{2j}$ are defined via the viscosities and the binary diffusion coefficient of each component in the gas mixture.
These values can be calculated from tabulated Lennard-Jones parameters of the idealized gas~\cite{welty2009fundamentals} and the local partial pressures and temperature of the gas mixture
\begin{equation}
\mu_{j} = 2.67\times10^{-6} \frac{\sqrt{M_j T}}{\sigma^2_j \Omega_{\mu}}
\end{equation}
and
\begin{equation}
D_{jk} = 1.86\times10^{-3}\;T^3\;\frac{\sqrt{\frac{1}{M_j} + \frac{1}{M_k}}}{P \sigma^2_{jk} \Omega_{D}}.
\end{equation}
The collision integrals $\Omega_D$ and $\Omega_{\mu}$ are a function of the Lennard-Jones parameters and the temperature. We evaluate those terms from a spline function based on the tabulated values from Appendix K in Ref.~\cite{welty2009fundamentals}.
To capture the full dynamics in a consistent way, the relaxation times need to be updated in each iteration and at each lattice site as the local conditions change over time.

The actual simulation runs in simulation units, which have to be properly related to SI units via the laws of the ideal gas together with the lattice spacing:
The characteristic velocity of an ideal gas of molar mass $M_j$ and temperature $T_0$ is given in SI units by
\begin{equation}
U = \sqrt{\frac{3 R T_0}{M_j}},
\end{equation}
where $R$ is the gas constant.
In internal LB units the characteristic velocity $U$ is equal to the lattice velocity $c_{ji}$. The lightest element is streamed on-lattice ($c_0 = 1$) and therefore sets the velocity scale~\cite{PhysRevE.89.063310}
\begin{equation}
U' = \frac{U(M_0)}{c_{0}}=\sqrt{\frac{3 R T_0}{M_0}}.
\end{equation}

The temperature scale in lattice units is defined with respect to a reference temperature, which is set to 1/3 in lattice units ($T'=T_{0}/(1/3)$).
The length scale is readily defined by the characteristic length in physical units $L$ and the discretized characteristic length in lattice units $L_{LB}$
\begin{equation}
L' = \frac{L}{L_{LB}}.
\end{equation}
The time scale is then calculated from the velocity and length scales as
\begin{equation}
t' = \frac{L'}{U'},
\end{equation}
so that the lightest species travels exactly one lattice constant in a single time step $\delta t = t'$. 
Conversion factors for the pressure, viscosity and binary diffusion coefficients can be derived in a similar way,
\begin{equation}
\mu' = L' U',  \;\;\; D' = L' U',  \;\;\; P'=U' U',\\
\;\;\;\kappa' = L' U', 
\end{equation}
while the density is set to have the same values as in SI units ($\rho'=1$). In most of the following simulations the lattice constant was set to be $5\cdot 10^{-3}~mm$, which for a reference temperature of $T = 403~K$ results in a time step $\delta t = 2.23\cdot10^{-9}~s$. The calculated relaxation times $\tau_1$ and $\tau_2$ are of the order $0.01\delta t - 0.9\delta t$.

\subsection{Interpolation}\label{sec:interpolation}
\begin{figure}
  \centerline{\includegraphics[width=8.5 cm]{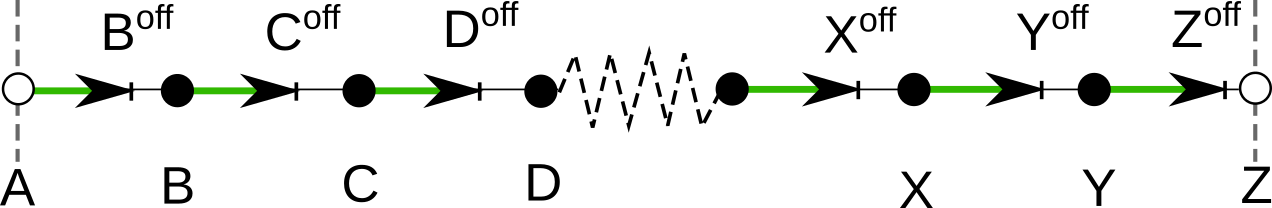}}
 \caption{\label{fig:interpol} Illustration of the streaming step for a heavier component with $c_j < 1$ mapping the population $g\big|^{on}_x \longrightarrow g\big|^{off}_{x+1}$ onto an off-lattice position, labelled by a superscript \textit{off}. A central difference interpolation scheme is used for all nodes except for wall nodes $A$ and $Z$, where a forward/backward scheme is used instead.
}
\end{figure}

In a general mixture, different species have different masses and thus also different microscopic velocities $c_{ji}$. This complicates the streaming step as the propagated distance within one time step is different for each species. Fig.~\ref{fig:interpol} illustrates the streaming for the velocities $c_{j}<1$, where the propagation distance during one time step is $\delta x = \lVert{c_{j}\rVert} \delta t < 1$. Starting from an on-lattice position (labelled with a capital letter), the populations $g\big|^{on}_x$ are streamed onto an off-lattice position $g\big|^{off}_{x+1}$.
Models employing multiple lattices (a lattice for every species) are possible. This, however, only shifts the difficulty into the collision step with widely unknown accuracy.

Here, we choose to work with a single lattice, which hosts the lightest gas species exactly. All heavier species are interpolated according to an upwind interpolation scheme suggested by Arcidiacono \textit{et al.}~\cite{PhysRevE.78.046711}, with the difference that here the transformed populations $g_{ji}$ (and not $f_{ji}$) are interpolated:
\begin{equation} \label{eq:interpol_general}
g_{ji}\big|_{x}^{on} = g_{ji}\big|_{x}^{off} + g_{ji}'\big|_{x}^{off} (1-\|{c_j}\|\delta t)
+ g_{ji}''\big|_{x}^{off} \frac{1}{2}(1-\|{c_j}\|\delta t)^2.
\end{equation}
The first  and second derivatives ($g_{ji}'$, $g_{ji}''$) are calculated from second-order central finite differences of neighboring off-lattice populations
\begin{equation} \label{eq:interpol_bulk} 
\begin{split}
g_{ji}'\big|_{x}^{off} &=  \frac{g_{ji}\big|_{x+1}^{off} - g_{ji}\big|_{x-1}^{off} }{2} \\
g_{ji}''\big|_{x}^{off} &=  g_{ji}\big|_{x+1}^{off} - 2\cdot g_{ji}\big|_{x}^{off} + g_{ji}\big|_{x-1}^{off}.
\end{split}
\end{equation}

Fig.~\ref{fig:interpol} illustrates the interpolation step for the streaming along lattice direction $i=1$ with the fluid domain between two walls positioned at lattice nodes $A$ and $Z$. For the bulk fluid domain Eq.~(\ref{eq:interpol_bulk}) is used to calculate the first and second derivatives. 
Close to the wall, however, e.g.~at lattice site $B$ where $A$ is a wall node and $A^{off}$ does not exist, the first and second derivatives are approximated through second-order forward finite differences as
\begin{equation} \label{eq:interpol_away}
\begin{split}
g_{ji}'\big|_{B}^{off} &=  \frac{4 g_{ji}\big|_{C}^{off} - g_{ji}\big|_{D}^{off} - 3 g_{ji}\big|_{B}^{off}}{2}, \\
g_{ji}''\big|_{B}^{off} &=  -2 g_{ji}\big|_{C}^{off} + g_{ji}\big|_{D}^{off} +  g_{ji}\big|_{B}^{off}.
\end{split}
\end{equation}
Respectively, interpolation onto a wall where $g_{ji}\big|_{x+1}^{off}$ does not exist, 
e.g. at the wall node $Z$ in Fig.~\ref{fig:interpol}, leads to 
\begin{equation} \label{eq:interpol_towards}
\begin{split}
g_{ji}'\big|_{Z}^{off} &=- \frac{4 g_{ji}\big|_{Y}^{off} - g_{ji}\big|_{X}^{off} - 3 g_{ji}\big|_{Z}^{off}}{2}, \\
g_{ji}''\big|_{Z}^{off} &=-2 g_{ji}\big|_{Y}^{off} + g_{ji}\big|_{X}^{off} + g_{ji}\big|_{Z}^{off}   .
\end{split}
\end{equation}
Note that the negative sign in the upper formula of Eq.~(\ref{eq:interpol_towards}) aligns the first derivative with the direction of interpolation, while the second derivative is symmetric.
Eqs.~(\ref{eq:interpol_away}) and (\ref{eq:interpol_towards}) are approximations to the exact derivatives, which induces an error to the overall mass conservation. This error correlates with the net flux perpendicular to the wall and with the density gradients. 
In long simulations with periodic boundary conditions around the simulation box this potentially becomes an issue. In practice, interpolated species can easily be replenished through suitable inlet boundary conditions, as exemplified in Sec.~\ref{sec:showcase}.

\subsection{Reactive boundary conditions}
\begin{figure}
\centerline{\includegraphics[width=0.5\columnwidth]{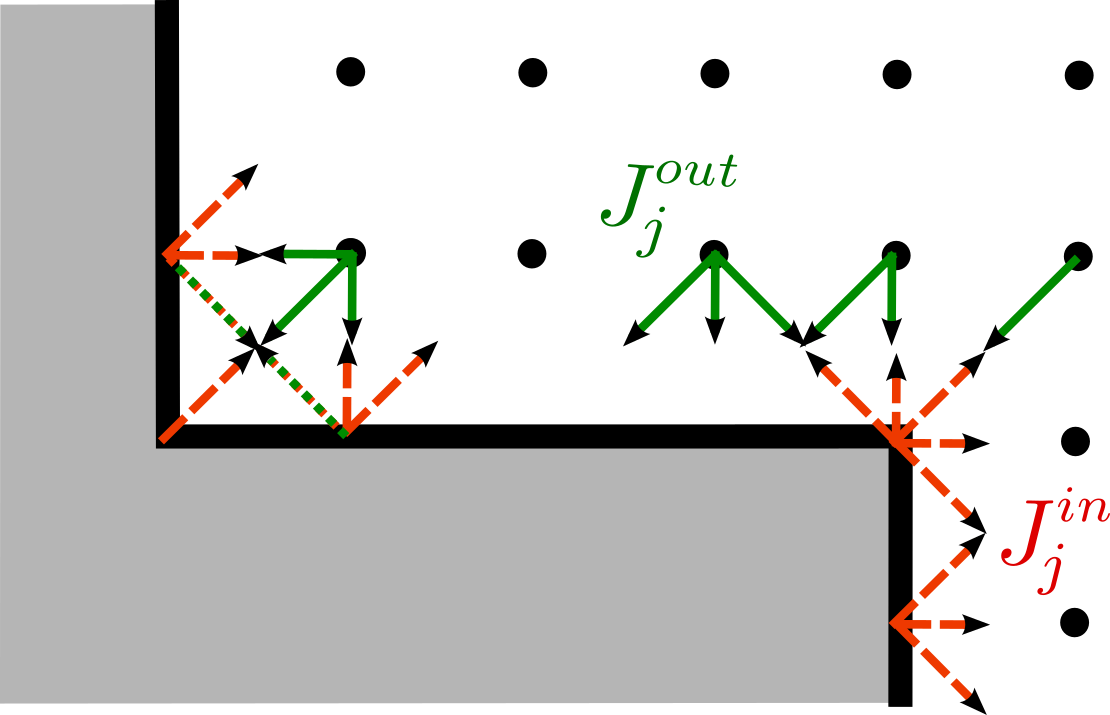}}
\caption{\label{fig:BC} Illustration of the boundary conditions at the wall: the populations flowing outwards (green) thermalize and react on the wall nodes (black line). The unknown populations flowing inward (red dashed) are calculated from Eq.~(\ref{eq:BC_gji}). Around concave corner sites populations act as outflowing and inflowing populations at the same time (red-green dotted).}
\end{figure}

Surface chemical reactions are simulated through appropriate boundary conditions at the gas-solid interface. This can be done by modifying the individual mass flux balance perpendicular to the wall by a reaction term $S_j$,
\begin{equation} \label{eq:mass_balance}
J^{out}_{j} - J^{in}_j = S_j ,
\end{equation}
where $J^{out}_{j}$ denotes the mass flux towards the wall. Respectively, $J^{in}_{j}$ is the mass flux pointing back into the gas domain. $J^{out}_{j}$ and $J^{in}_{j}$ can be expressed by the distribution function as 
\begin{equation} \label{eq:J_out}
J^{out}_{j} = \sum_{\{i | c_{ji\alpha} n_\alpha < 0\}} |c_{ji\alpha}n_\alpha | f_{ji}, 
\end{equation}
where the sum is restricted to values of $i$ such that the projection of $c_{ji\alpha}$ onto the surface normal $n_\alpha$ is negative (see Figure~\ref{fig:BC}). Analogously,
\begin{equation} \label{eq:J_in}
J^{in}_{j} =\sum_{\{i | c_{ji\alpha} n_\alpha > 0\}} |c_{ji\alpha}n_\alpha |f_{ji}.
\end{equation}

Various schemes exist in the literature to incorporate boundary conditions on the basis of the mass flux. Many of these schemes generalize athermal boundary conditions e.g. the bounce back scheme \cite{kang2007improved,chen2013pore,PhysRevE.85.016701}. 
In many applications it is desirable to simulate a heated catalyst surface, i.e.~to include a wall temperature $T_{w}$ as a parameter of the simulation. One elegant way to achieve this flexibility is to adopt diffusive boundary conditions~\cite{sofonea2005boundary,ansumali2006entropic,PhysRevE.78.046711,hecht2010implementation}. Diffusive boundary conditions assume that populations at the wall have enough time to thermalize to equilibrium defined by the wall temperature and can thus be written as
\begin{equation} \label{eq:BC_g_ji_1}
f_{ji} =  f^{w}_{ji} (\rho_{w}, U_{w}, T_{w}) = \rho_{w} f^{eq}_{ji} (\mathbbm{1}, U_{w}, T_{w}) ,
\end{equation}
where $\rho_{w}$ is the density at the wall, and $U_{w}$ denotes a possible slip velocity at the wall. By further assuming that the correction terms ($\Phi$ and $\Psi$) can be neglected at the wall,
the populations $f_{ji}$ can be replaced by the transformed $g_{ji}$ in Eq.~(\ref{eq:J_out}) and Eq.~(\ref{eq:J_in}).
Together with Eq.~(\ref{eq:mass_balance}), the unknown inflowing populations are then given by
\begin{equation} \label{eq:BC_gji}
g_{ji} =  f^{w}_{ji}  \frac{\sum_{\{i | c_{ji\alpha} n_\alpha < 0\}} |c_{ji\alpha}n_\alpha | g_{ji} + S_j}{\sum_{\{i | c_{ji\alpha} n_\alpha > 0\}} |c_{ji\alpha}n_\alpha | f^{w}_{ji}},
\end{equation}
as derived by Arcidiacono et al.~\cite{PhysRevE.78.046711}.

Boundary conditions are implemented such that the wall is positioned on the solid node (black line in Figure~\ref{fig:BC}).
The reaction term $S_j$ is a function of the local density at the wall $\rho_{w}$. Similar to the procedure described in Ref.~\cite{PhysRevE.78.046711} $\rho_{w}$ is determined iteratively: initially, $\rho_{w}$ is set to the value of the first fluid node. By evaluating the reaction term with the help of Eq.~(\ref{eq:mass_balance}), the unknown populations in Eq.~(\ref{eq:BC_gji}) can be computed. The result is then used to evaluate a new density by solving Eq.~(\ref{eq:BC_g_ji_1}) for $\rho_w$.
This process is iterated until $\rho_w$ is converged (within $10^{-8}$).
With the converged value of $\rho_w$, the final unknown populations at the wall are computed from Eq.~(\ref{eq:BC_gji}) and stored for further operations. 

It is also possible to mimic a wall positioned half-way between the first fluid node and the wall node. This, however, requires an additional interpolation step as described in Ref.~\cite{PhysRevE.78.046711}.

Applying the boundary conditions to arbitrary geometries consisting of plane wall nodes and concave and convex corners is in principle straightforward. The calculated outflowing and inflowing mass flux thereby only differ by the number of populations entering Eq.~(\ref{eq:BC_gji}), as illustrated in Figure~\ref{fig:BC}.
We, however, need to point out that the interpolation of the streaming step around concave corner sites is problematic. For populations streaming from wall to wall (red-green dotted arrows in Fig.~\ref{fig:BC}) neighboring populations in the streaming direction are missing on both sides. This prohibits any interpolation which in turn leads to artificially large mass fluxes in these directions.
In this work we restrict the surface geometry to be constructed from planar and convex corner nodes, where such issues do not occur.

The surface chemical reaction not only changes the chemical composition, but in general also the temperature of the system according to the reaction enthalpy $\Delta H_r$,
\begin{equation} \label{eq:enthalpy_T}
\Delta T = \frac{\Delta H_r}{c_p} ,
\end{equation}
where $c_p$ is the isobaric heat capacity.

Previous thermal LBM approaches use the concept of isothermal walls~\cite{PhysRevE.89.063310}, which reflects adiabatic coupling to an infinite bath or solid nodes with infinite heat capacity.
For many applications involving chemical reactions it is desirable to capture heat accumulation at the reactor surface and in the reactor volume. This requires to solve the heat equation in the solid nodes, together with a coupling between the solid and the fluid domains. In Sec.~\ref{sec:fd-lbm} we suggest a simple and efficient way to do that.

\subsection{Benchmarking mass and thermal diffusion}
To show the quality of the corrected model the mass diffusivity, thermal diffusivity and the viscous heat dissipation of the LB simulation are individually benchmarked in three different setups.
All simulations are performed with a gas mixture of $\mathrm{H_2}$, $\mathrm{H_2O}$, $\mathrm{CO}$ \& $\mathrm{CO_2}$. This yields a maximum ratio of molar masses of 22, which scrutinizes the robustness of the interpolation scheme described in the previous section. If not stated otherwise, the length of a lattice cell corresponds to $5\cdot10^{-3}~mm$.

\begin{figure}
  \centerline{\includegraphics[width=0.6\columnwidth]{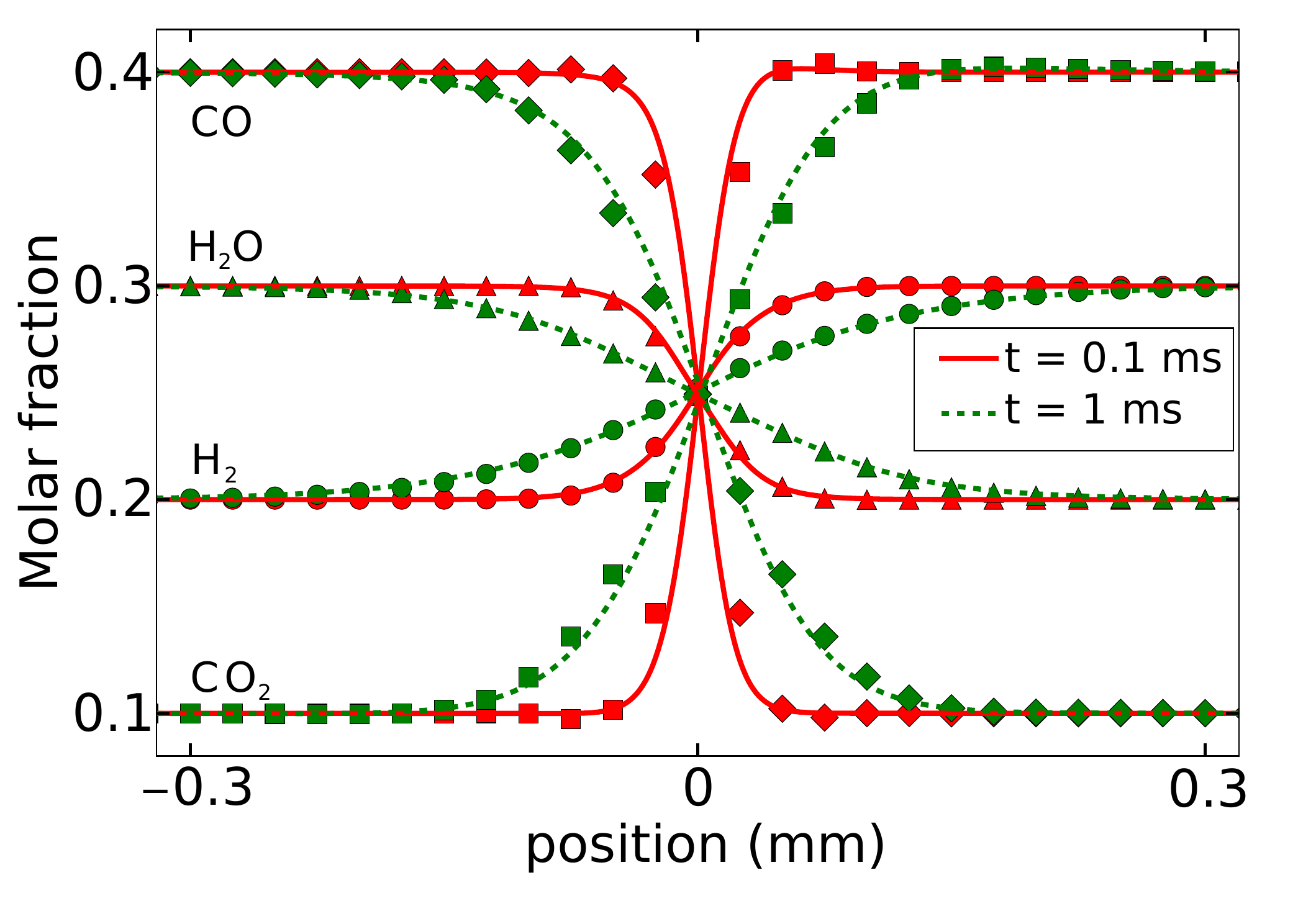}}
 \caption{\label{fig:mass_diffusion} Mass diffusion of the multicomponent mixture: comparison of molar fractions of the four different species from the LB simulation (symbols) 
 with the reference solution (lines), at time $t$=0.1 ms (green) and $t$=1 ms (red).}
\end{figure}

To test the correct mass diffusion a quasi infinite symmetric one-dimensio\-nal system is set up with two zones having adjugate initial molar fraction, the molar fractions of $\mathrm{CO_2}$,  $\mathrm{H_2}$,  $\mathrm{H_2O}$ and  $\mathrm{CO}$  are $0.1, 0.2, 0.3$ and $0.4$ for $x<0$ and, respectively $0.4, 0.3, 0.3,$ and $0.1$  for $x>0$, at time $t=0$, as indicated in Fig.~\ref{fig:mass_diffusion}. The system consists of $2000\times 1$ lattice sites with a reference temperature of $293~K$.

The gradients of species concentrations lead to diffusion fluxes combined with small convection fluxes caused by pressure drops that are originating from the differences in the speed of diffusion of the components. All these effects are not to be described analytically and need to be modelled by other established methods. We compare the diffusion profiles obtained with our LB simulation to the ones obtained with a commercial FEM software (COMSOL Multiphysics) with the same system parameters, binary diffusion coefficients and initial values. 

As can be seen from Fig.~\ref{fig:mass_diffusion}, the simulation results show very good agreement with the reference simulation for all four chemical species.

The macroscopic heat transport of the gas mixture is tested by simulating the heat conduction between two isothermal infinite plates with different temperatures of $T_c =293~K$ and $T_h =T_c + \Delta T =303~K$.
Between the two plates a gas mixture of $\mathrm{H_2}$, $\mathrm{H_2O}$, $\mathrm{CO}$ \& $\mathrm{CO_2}$ with the molar fractions of 20\%, 30\%, 10\% and 40\% at $1~bar$ and $293~K$ is placed.
\begin{figure}
\centerline{\includegraphics[width=0.6\columnwidth]{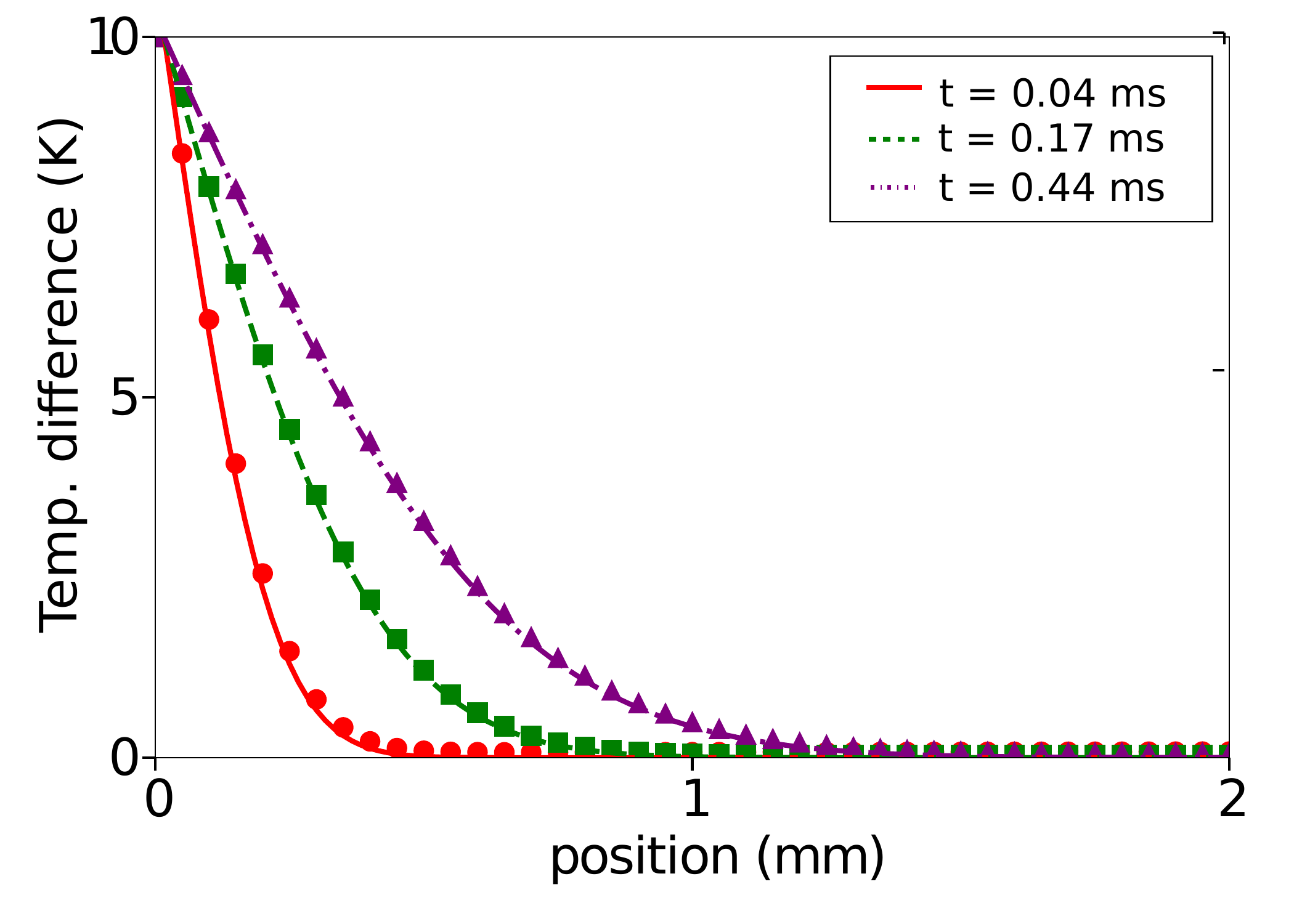}}
 \caption{\label{fig:heat_diffusion} Thermal diffusivity in the multicomponent mixture between two parallel plates: comparison of the temperature profile of the gas mixture from the LB simulation (symbols) 
 with the analytical solution (lines), at different time steps. The simulation details are given in the text.}
\end{figure}
The system is discretized by 300 lattice sites between the plates, resembling a distance of $L=5~mm$. Together with a Prandtl number $Pr_j=0.7$ for each species, the mixture averaged thermal diffusivity ($\alpha=\kappa/2C$) is calculated as $\alpha=2.8~cm^2/s$.

Diffusive boundary conditions~\cite{sofonea2005boundary,ansumali2006entropic} are applied to both walls in the x-direction, and periodic boundary conditions in the y-direction, respectively.
The resulting temperature profiles (Fig.~\ref{fig:heat_diffusion}) show excellent agreement with the analytical solution of the heat equation given by
\begin{equation}
T(x,t) = T_0 + \Delta T \bigg[1- erf \bigg(\frac{x}{2\sqrt{\alpha t}}\bigg)\bigg].
\end{equation}

Finally, the viscous heat dissipation is tested in terms of a thermal Couette flow. Similar to the previous test case, the gas mixture is placed between two parallel walls, but now with the warmer wall moving tangential to the fluid domain with a wall velocity $U_{w} =224~m/s$ and with a small temperature difference of $\Delta T = 1~K$ between the two plates. The resolution used in this simulation is $10^{-3}~mm$.
Again, diffusive boundary conditions are applied to both plates, and periodic boundary conditions are applied in the flow direction.

The Navier-Stokes equations predict the steady-state temperature profile to be a function of the Prandtl number as
\begin{equation}
\label{eq:thermal_couette}
T(x) = \Delta T \frac{x}{L} +  \frac{Pr U^2_{w}}{2 c_p \Delta T } \frac{x}{L}\bigg(1-\frac{x}{L}\bigg) ,
\end{equation}
with $c_p$ being the isobaric heat capacity of the ideal gas in two dimensions.

\begin{figure}
\centerline{\includegraphics[width=0.6\columnwidth]{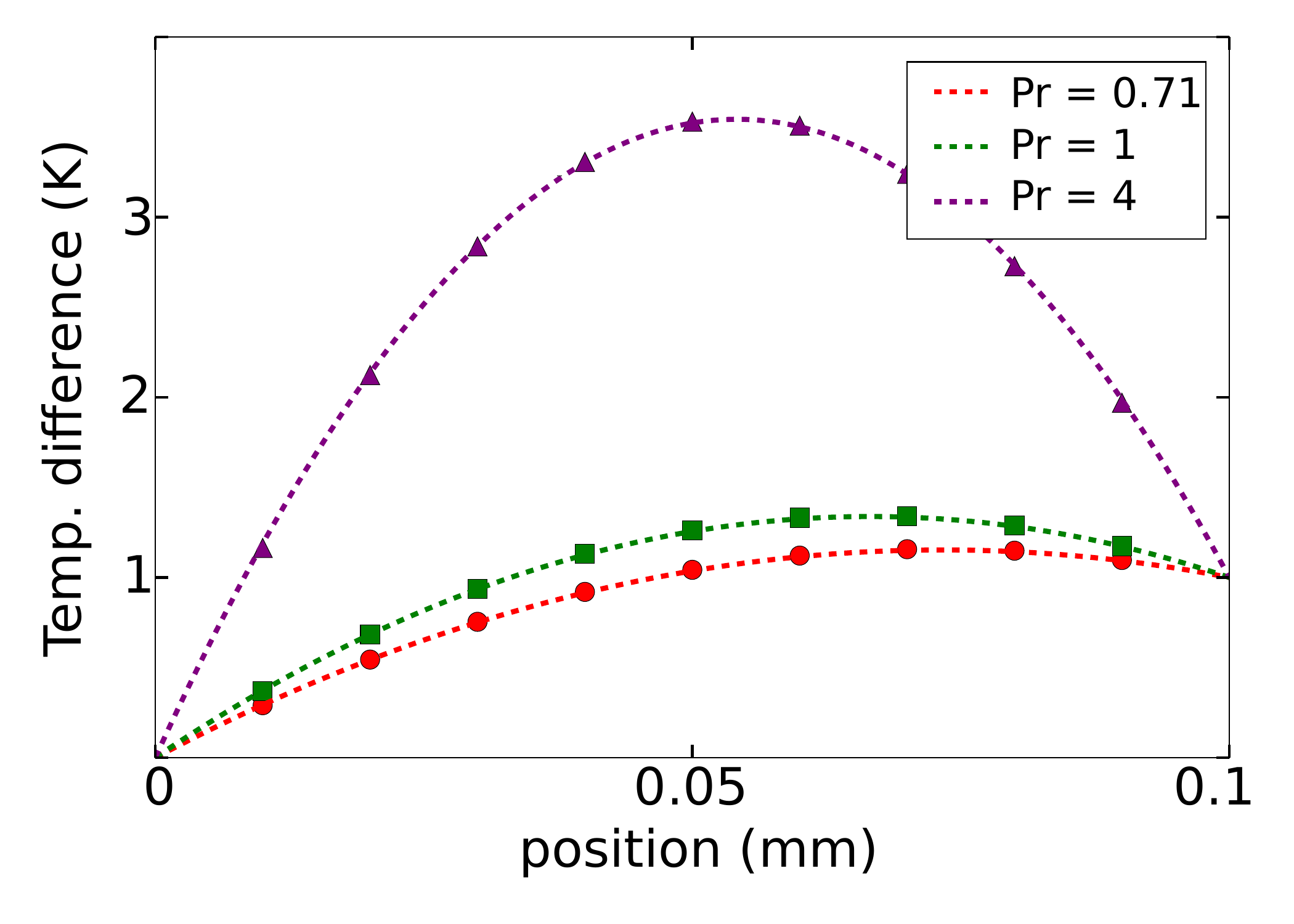}}
 \caption{\label{fig:thermal_couette} Thermal Couette flow in a mixture of 0.001\% H$_2$, 49.999\% $\mathrm{H_2O}$, 49.999\% $\mathrm{CO}$ and 0.001\% $\mathrm{CO_2}$ at different Prandtl numbers: comparison of the steady-state temperature profile resulting from the LB simulation (symbols) and the analytical solution (lines). Viscosity and thermal conductivity of the individual components is calculated consistently from the Lennard-Jones parameters. The Prandtl number is adjusted with the help of Eq.~(\ref{eq:forcing_correction_Pr}). }
\end{figure}
Fig.~\ref{fig:thermal_couette} shows the steady-state temperatures (after $5\times10^{5}$ time steps) for three different Prandtl numbers ($Pr=0.71, 1, 4$). $Pr = 0.71$ is a typical value for a gas at ambient conditions, $Pr=1$ reflects the case when no Prandtl number correction is applied ($\Phi^{Pr}_{ji} = 0$), and $Pr=4$ is the case where no energy correction at all is applied ($\Phi^{Pr}_{ji} = 0$ \& $\Phi_{ji} = 0$). The agreement with the analytical solution (Eq.(\ref{eq:thermal_couette})) is excellent over the whole simulation domain, which shows the quality of the overall model.

\section{Heat conduction across a solid-fluid interface}
\label{sec:fd-lbm}
In order to accurately capture effects like heat accumulation two ingredients are needed: (i) the thermal conduction in the solid and (ii) coupling of the temperature field across the solid-fluid interface.
The thermal conduction within the solid is described by the heat equation 
\begin{equation}\label{eq:heat_equation}
\partial_t T =  \alpha_{s} \nabla^2 T  ,
\end{equation}
with $\alpha_{s}$ being the thermal diffusivity within the solid.
In this paper the heat equation is solved with a second-order forward Euler finite difference (FD) scheme. To avoid additional computational overhead for interpolation the identical real-space lattice and time integration step $\delta t$ as for the LB simulation are used.
The time propagation along each lattice direction is evaluated in each LB iteration as
\begin{equation}\label{eq:heat_equation_fd}
\begin{split}
T^s_{x}(t+ \delta t) = T^s_{x}(t) + \alpha_{s} {T^s}^{''}_{x}(t) ,
\end{split}
\end{equation}
where the second derivative 
\begin{equation}\label{eq:heat_equation_fd_2nd_deriv}
\begin{split}
{T^s}^{''}_{x}(t) =&   T^s_{x-1}(t) - 2 T^s_{x}(t) + T^s_{x+1}(t),
\end{split}
\end{equation}
is calculated from central finite differences of the solid temperature $T^s$ for each solid node $x$.

For $\alpha_{s} > 0.5$ the stability of the forward Euler scheme is challenged. For such values a fully explicit scheme, e.g.~Crank-Nicholson, is recommended. Extending the current implementation into this direction is straightforward. In this study, however, we focus on materials like porous silicate, in which $\alpha_{s}$ is of the same order as the thermal diffusivity in the gas domain. The stability of the overall scheme is therefore governed by the stability of the LBM part, i.e. the range of values for $\tau_1$. Or in other words: In this combined FD-LBM scheme the LBM time step is very small anyway, such that the FD part is stable.

Using the same lattice for the FD as well as for the LBM calculation allows for a simple definition of Dirichlet-like boundary conditions in each iteration:
\begin{equation}\label{eq:wall_temperature}
T_{s}(t+ \delta t) = T^s(t)
+ \alpha^s(T^s_{x-1}(t) -  T^s_{x}(t))
- \alpha^f\frac{\rho^{f} c_{p }^f }{\rho^{s} c_{p}^s }(T^f_{x+1}(t) -  T^s_{x}(t)),
\end{equation}
where the second and third terms account for the heat transfer toward the solid and and fluid nodes next to the boundary, respectively. 
Coupling of the solid temperature to the temperature field in the fluid domain is realized through the diffusive boundary conditions (Eq.~(\ref{eq:BC_gji})), where the wall temperature $T_{w}$ is equal to the temperature of the first solid node $T^s$. The convective term $\textbf{v}\cdot\nabla T$ in Eq.~(\ref{eq:wall_temperature}) can be added for the case of a moving and/or slip boundary and is skipped here for simplicity.

\begin{figure}
\centerline{\includegraphics[width=0.6\columnwidth]{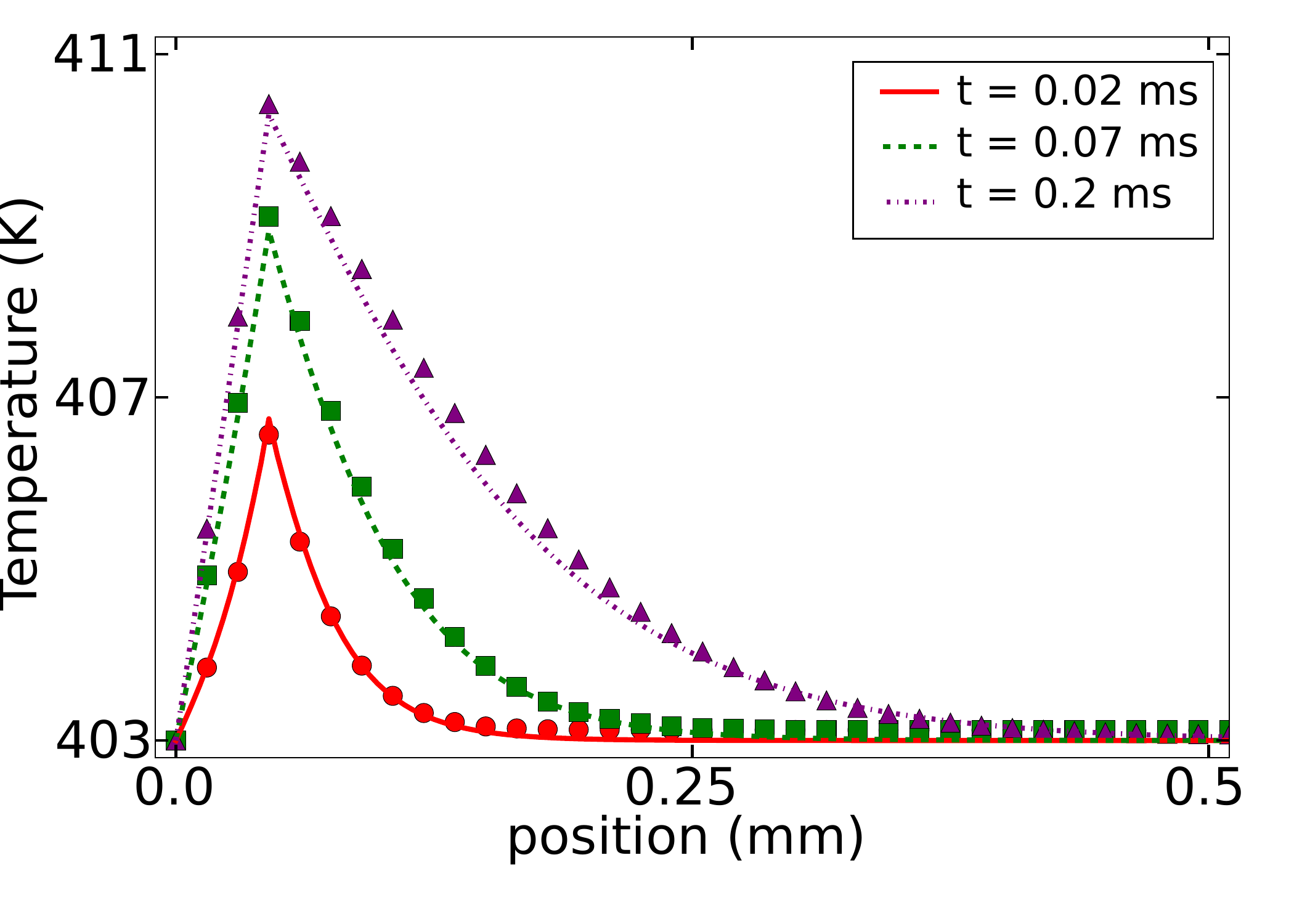}}
\caption{\label{fig:temp_time_dep} Time dependent temperature profiles in a quasi-1D channel with reactive walls. Temperature distributions obtained with the LBM (symbols) are compared to the ones calculated by COMSOL (lines). The position of the solid-fluid interface is placed at $0.05~mm$.
}
\end{figure}

Two systems are exploited to benchmark the solid-fluid thermal coupling and the results are compared to analogous systems simulated by COMSOL Multiphysics. First, a quasi-1D  channel with two solid walls and reactive boundary conditions at the interfaces between solid and fluid is modeled in a transient regime. Reactive boundary conditions at the interface catalyze the exothermic water-gas-shift reaction ($\mathrm{CO + H_2O \longrightarrow H_2 + CO_2}$) with reaction enthalpy $20.5~kJ/mol$ and rate prefactor $127\cdot10^3~m/s$ . 
The corresponding heat produced is the reaction enthalpy $\Delta H_r$ times the molar change of the reactant, e.g. $\mathrm{CO}$. The outer boundaries of the solid are kept at constant temperature. The solid density $\rho^s$, heat capacity $c_p^s$ and thermal diffusivity $\alpha^s$ are, correspondingly, $1000~kg/m^3$, $2~kJ/kg~K$ and $3.4~cm^2/s$. The resulting temperature distributions at three moments in time are shown in Fig.~\ref{fig:temp_time_dep}. A small discrepancy of the temperature from the values calculated with the heat equation (COMSOL) is visible in the fluid domain. Since in the LBM the fluid is weakly compressible, a temperature gradient always induces a pressure wave. This pressure wave transports thermal energy at the speed of sound, which is a mechanism not captured by Eq.~(\ref{eq:heat_equation}).
Besides the correct thermal diffusion, any temperature gradient at the interface heats the fluid through this mechanism as well. In the reference calculation (COMSOL), on the other hand, only the heat diffusion equation together with the \textit{Transport of Concentrated Species} interface  were used, since including the solution of the Navier-Stokes equation caused stability issues. This way, no heat transfer induced by pressure wave effects can be captured.

\begin{figure}
\centerline{\includegraphics[width=0.6\columnwidth]{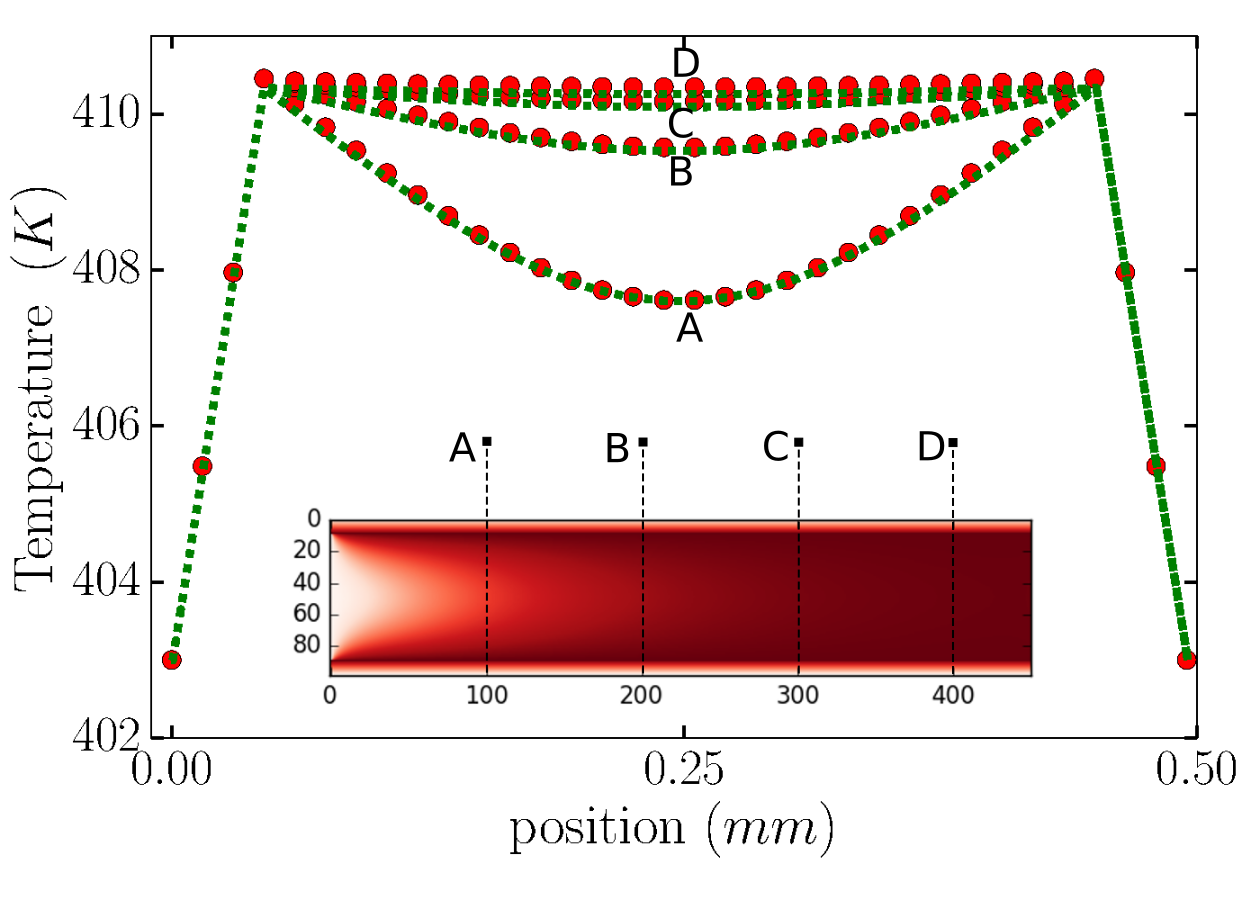}}
\caption{\label{fig:temp_steady_state} Steady state temperature profiles in a 2D channel with reactive walls. Temperature distributions obtained with LB (symbols) are compared to the ones calculated by COMSOL (lines). The cross-section line positions at which the temperature profiles are plotted are indicated on the inset.
}
\end{figure}

The second system is a 2D channel (100 by 500 lattice sites) with reactive walls and a steady-state flow. The reaction enthalpy and rate are the same as in the simulation above. In Fig.~\ref{fig:temp_steady_state}, the inset shows the channel temperature distribution, including solid walls with reactive boundaries. The mixture of gases flows in through the inlet (left) with a constant temperature of $403~K$. The cross lines (A-D) at which the temperature is plotted in the main graph (red symbols) are shown as well. The green lines depict the temperature profiles obtained with COMSOL using the module {\it Conjugate Heat Transfer for Laminar Flow}, coupled to {\it Transport of Concentrated Species and Laminar Flow} interfaces. A very good match of the temperature profiles both within the fluid and at the solid-fluid boundary is obtained along the channel.

The quality of the overall thermal coupling through the interface (incl. Eq.~(\ref{eq:heat_equation_fd}) and  Eq.~(\ref{eq:BC_gji}) with Eq.~(\ref{eq:wall_temperature})) is also scrutinized in a quasi 1-dimensional test system (with $200 \times 3$ lattice sites) consisting of a solid and a fluid domains. Periodic boundary conditions are applied in all four directions, hence comprising two solid-fluid interfaces.
Reactive boundary conditions at the interface catalyze the exothermic water-gas-shift reaction as in previous tests. 
The corresponding change of heat is the reaction enthalpy $\Delta H_r$ times the molar change of the reactant, e.g. $\mathrm{CO}$.
The response of the system is measured by calculating the change of thermal energy in the fluid and the solid domain, which allows to define the error in enthalpy conservation as
\begin{equation}\label{eq:error_coupling}
\varepsilon = \sum \frac{ c_p(T  -  T_0  )- \Delta H_r \cdot \Delta C_{CO}  } 
{\Delta H_r\cdot\Delta C_{CO}  },
\end{equation}
where $T_0$ is the starting temperature.
The error $\varepsilon$
is measured and found to be of the order $10^{-5}$ for a wide range of relaxation times $\tau_1$. 

We test the generality and transferabilty of the coupling scheme employing a variety of simulation domain configurations over a wide range of sensible values for applications in chemical reactors. We find that the dependence on the system size (if larger than $N=20$) and the solid fraction is negligible. The observed error is also independent of the initial temperature $T_0$ and independent of the reaction rate and the reaction enthalpy. We also measure that the enthalpy uptake of the solid and the fluid fraction is in very good agreement with the respective heat capacities.

\section{Showcase}
\label{sec:showcase}
As a showcase for the overall model we present simulation results of a model fixed-bed reactor consisting of three cubes with quasi-random distribution in a channel. The simulation domain is discretized on a lattice with $N_x \times N_y = 500 \times 200$ lattice nodes, assuming a resolution of $5~\mu m$ per lattice unit. The edge length of the cubes is as low as $200~\mu m$ showing the applicabilty of the scheme for micro-porous material.
The surface of the cubes catalyses the water-gas-shift reaction, involving 
a gas species with very different viscosities, diffusivities and masses, rendering this a highly diagnostic set-up to scrutinize the scheme.
Conditions at the inlet (left side) are fixed to ensure a continuous flow of constant temperature ($403~K$), constant pressure ($1~bar$), Poiseuille-like flow velocity ($\mathrm{u_{x-max} = 24}~m/s$) in positive x-direction and a constant molar fraction of 49.9999\% of the CO and $\mathrm{H_2O}$ and 0.0001~\% of $\mathrm{H_2}$ and $\mathrm{CO_2}$. This can be achieved by setting the distributions at the inlet nodes to equilibrium calculated with these values.
Diffusive boundary conditions are applied on the channel walls.
The dynamic viscosities (thermal conductivities) and the binary diffusion coefficients of each species are calculated from the Lennard-Jones parameters of the gas, at given local conditions. 
For the material properties of the solid nodes we use a density $\rho^s=1000~kg/m^3$ and specific heat of $c_p^s=2.~kJ/kg~K$.
The thermal diffusivity within the solid cubes is set to $\alpha^s = 0.03$ in LB units. The temperature of the outer boundary of the channel walls is kept constant at $403~K$.

The reaction term $S_j$ is calculated from the rate constant $k$ given by the Arrhenius equation ($k=A\exp{[-E_a/k_bT_{w}]}$) at the given local wall temperature, the reactant concentrations at the wall and the molar masses as
\begin{equation}\label{eq:showcase_Sj}
S_j =  \pm M_j k\cdot C_{CO}C_{H_2O} \cdot \sqrt{\frac{M_0}{M_j}}.
\end{equation}
The minus sign is used for reactants and the plus sign for product species.
Note that the last factor stems from the fact that $S_j$ corresponds to a mass flux and not a mass density. It therefore requires the same scaling as the microscopic velocities $c_j$. We use an activation energy $E_a$ of $84~kJ/mol$ taken from literature \cite{B912688K}.
For the purpose of a showcase, we use a value for the rate prefactor $127000~m/s$ and the reaction enthalpy $\Delta H_r = -4.1~kJ/mol$.

\begin{figure}
\centerline{\includegraphics[width=0.7\columnwidth]{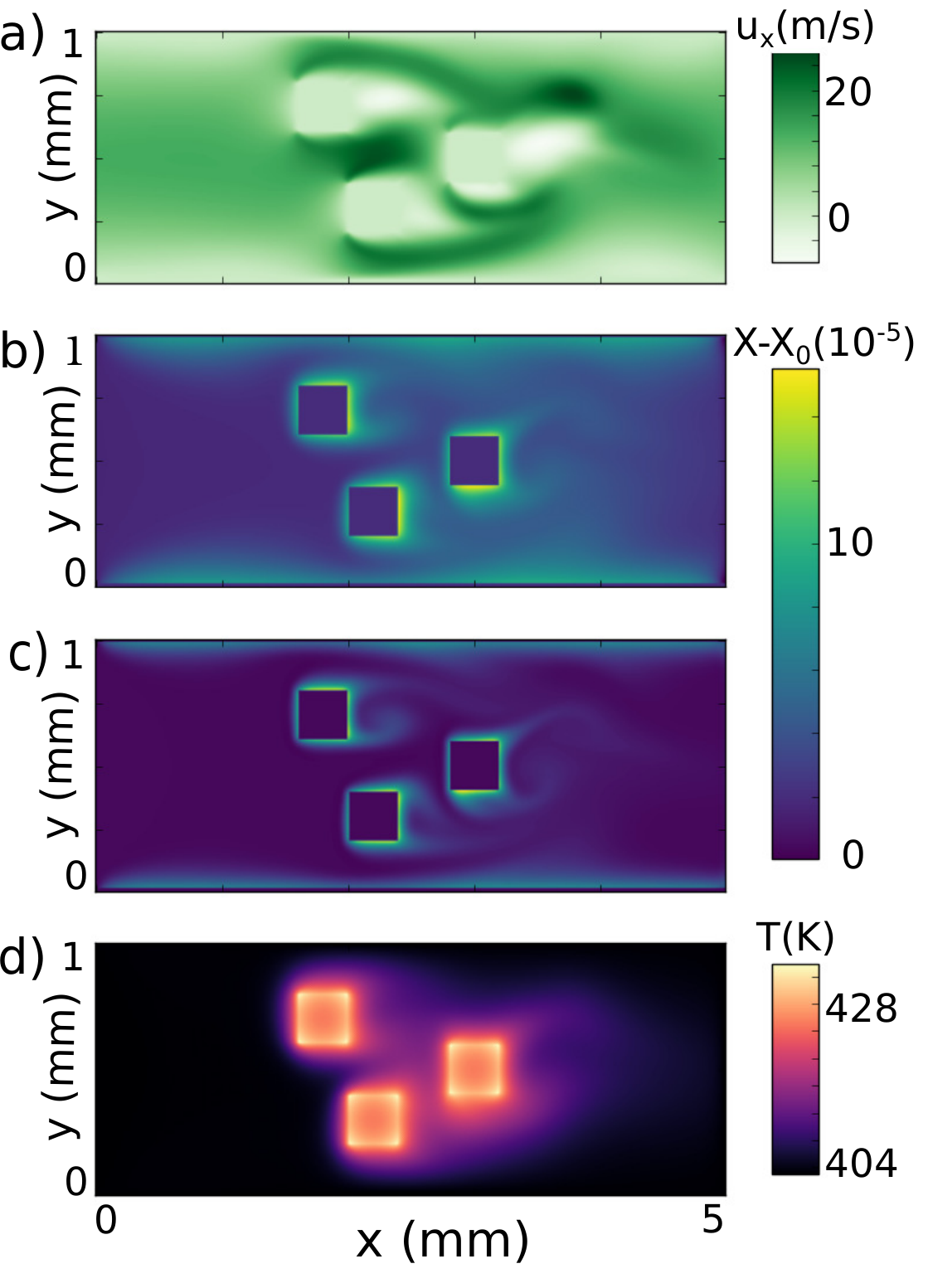}}
\caption{\label{fig:showcase} Reactive flow through a model porous network. The contour maps show the velocity field (a), molar fractions of the products $\mathrm{H_2}$ (b) and $\mathrm{CO_2}$ (c) and the temperature profile (d) after 50000 time steps. The surface of the cubes catalyzes the water-gas-shift reaction. The inlet is on the left hand side and the outlet on the right, diffusive boundary conditions are applied on the walls.
}
\end{figure}

The given inlet and outlet boundary conditions together with no-slip walls lead to a flow profile as shown in Fig.~\ref{fig:showcase} (a). The solid cubes constrain the gas flow through a network of small channels (pores), which locally increases the flow velocity. Instabilities of the flow field can be seen at this $Re$ number of about $330$. The effect of the reactive boundary conditions can be seen in Fig.~\ref{fig:showcase} (b) and (c). The molar fractions of the two product species $\mathrm{H_2}$ and $\mathrm{CO_2}$ increase at the surface of the solid cubes, and are then transported away by the gas stream. The product concentration is highest on the backside of the cubes. There, the flow velocity is low, which leads to an accumulation of the product species. This effect is much more pronounced for the heavier $\mathrm{CO_2}$. The amount of both product species, i.e.~the concentrations integrated over the entire simulation domain, are equal to within 1\%, which shows the quality of mass conservation even for large mass ratios. The local concentrations, however, deviate significantly, which stems from the fact that the diffusion coefficient of $\mathrm{H_2}$ is larger by a factor of five as compared to $\mathrm{CO_2}$. Only through the individual treatment of each species with their individual transport parameters and accurate correction terms, we can observe and study these local accumulation effects.

A similar complexity is also visible in the temperature field as shown in the Fig.~\ref{fig:showcase} (d). Along with the change of chemical composition the reaction at the surface releases reaction enthalpy, which heats up the surface of the solid cubes. On the one hand, the heat is conducted into the solid domain, and raises the temperature there. On the other hand, the heat diffuses into the fluid domain and is transported away by the flow field.
While there is a general increase in temperature in flow direction (from left to right), the detailed features on the surface of the cubes are rich. The highest temperature regions in the fluid agree to those with highest product concentration, namely the backside of the cubes. Here, the flow velocities are smallest which in turn leads to diffusion limited heat transport. The corners of the cubes are, in turn, the hottest parts of the solid. This stems from the lack of heat transport paths away from the corners. A very small part of the heat produced is transported into the fluid because of its comparatively small density and heat capacity (according to equation~\ref{eq:wall_temperature}).  
The fact that the temperature significantly decreases towards the outlet shows that the snapshots in Fig.~\ref{fig:showcase} (after 50000 time steps) do not correspond to steady state conditions. Of course, also the full dynamics, such as the dynamical temperature-reaction coupling, are accessible through the proposed method. Such effects however happen on time scales, which require considerable computational time going beyond the mere purpose of a showcase. 

We stress that the temperature increase is to the vast extend due to the reaction enthalpy. However, local temperature variations of up to 0.4~K also appear due the dynamic pressure variation around the obstacles. While this effect is to the largest extend physically justified it also leads to minor artifacts around the corner sides of the cubes, which stems from the large flow velocities and therefore large gradients around the corner. 
These artifacts can be suppressed by decreasing the flow velocity or through the refinemenmt of the computational grid. 

\section{Conclusions}
\label{sec:conclusion}
We presented a hybrid finite-difference LBM scheme for the study of enthalphy consistent catalytic flows through porous media. 
The gas phase domain is decribed by a multicomponent thermal LBM following the model by Kang \text{et al.}~\cite{PhysRevE.89.063310}. Reactive boundary conditions are used to simulate a chemical reaction at the surface of an obstacle in a very flexible way. The reaction enthalpy consistently heats up the fluid as well as the solid domain, where the heat equation is solved using a finite-differences algorithm. We showed that the proposed coupling mechanism for the enthalpy is very accurate.
In a final showcase we exemplified the application in the form of a catalytic flow through a model porous material, revealing rich information of the local temperature and molar fraction distributions.

In applications resembling real systems with defined system size, temperature, pressure and flow velocity a reasonable value for the grid spacing $\delta x$ in SI units has to be chosen. 
Together with the velocity scale, which is defined by the temperature and molar mass of the lightest element, the grid spacing also defines the time step $\delta t$ in SI units. 
Small length scales thereby imply small time steps. 
For the showcase example in Sec.~\ref{sec:showcase} the grid resolution of $\mathrm{\delta x = 5\mu}$ yields a time step of $\delta t = 2~ns$.
This of course encumbers the applicability when slow processes are to be observed on small length scales. On the other hand, fast processes require a reasonable time resolution which requires a high grid resolution.
Modern supercomputers and their ever-increasing computational power, however, allow to access long time scales in full resolution nonetheless.
To this end the fact that the proposed algorithm is straightforward to implement for massively-parallel infrastructure is a key advantage.

\section{Acknowledgments}

We thank Nikolaos I. Prasianakis, David M.
Smith, Marco Haumann and Peter Wasserscheid for fruitful
discussions. The authors acknowledge financial support
by the Deutsche Forschungsgemeinschaft (DFG)
within the Cluster of Excellence "Engineering of Advanced
Materials” (project EXC 315) (Bridge Funding) and CRC1411 "Design of Particulate Products". We further acknowledge support by
the Bundesministerium f\"ur Bildung und Forschung
BMBF within project “Tubulyze” (project number 03SF0564E).


\end{document}